\documentclass[prd,aps,preprint,tightenlines,nofootinbib,superscriptaddress]{revtex4}

\usepackage{epsfig}
\usepackage{bm}
\usepackage{amssymb}
\usepackage{amsmath}
\usepackage{color}

\usepackage{color}
\usepackage{babel}
\usepackage{float}
\usepackage{mathrsfs}
\usepackage{amsmath}
\usepackage{cancel}
\usepackage{graphicx}
\usepackage{setspace}

\usepackage{slashed}
\usepackage{simplewick}
\usepackage{ulem}

\begin{document} 

\author{Bao-dong~Sun}
 \affiliation{\normalsize\it Key Laboratory of
Particle Physics and Particle Irradiation (MOE),Institute of
Frontier and Interdisciplinary Science, Shandong University,
(QingDao), Shandong 266237, China }

\author{Ze-hao~Sun}
 \affiliation{\normalsize\it Key Laboratory of
Particle Physics and Particle Irradiation (MOE),Institute of
Frontier and Interdisciplinary Science, Shandong University,
(QingDao), Shandong 266237, China }

\author{Jian~Zhou}
 \affiliation{\normalsize\it Key Laboratory of
Particle Physics and Particle Irradiation (MOE),Institute of
Frontier and Interdisciplinary Science, Shandong University,
(QingDao), Shandong 266237, China }

\title{Trace anomaly contribution to hydrogen atom mass   }

\begin{abstract}
We compute  trace anomaly contribution to hydrogen atom mass, which turns out to be related to the part of the Lamb shift. This finding might shed new light on our understandings of the mass structure  of QCD bound states, such as, proton. 
\end{abstract}
\maketitle

The origin of  proton mass is one of the most fundamental questions remain to be answered   in hadronic physics study.  The  current light quark masses generated through the Brout-Englert-Higgs mechanism only makes up  a small part of proton mass.   At the classical level, the proton mass  vanishes identically in the massless limit due to the exact cancellation between the quark's kinetic energy and the negative potential energy according to the relativistic virial theorem~\cite{Brack:1983ht},  the field theory formulation of which  states that  the trace of the QCD energy momentum tensor(EMT) vanishes in the chiral limit. Therefore, the large piece of  proton mass essentially originates from the quantum effect, i.e. the trace anomaly of the energy momentum tensor~\citep{Adler:1976zt,Nielsen:1977sy,Collins:1976yq}, which comes as the consequence of the violation of the approximate conformal symmetry of QCD~\cite{Collins:1976yq}.

Due to the non-perturbative nature of low energy QCD, it appears to be impossible to solve the problem analytically. Instead, one can try to gain some insights into the proton mass structure by decomposing it into different pieces~\citep{Ji:1994av,Ji:1995sv} based on the QCD EMT. These different  contributions  can be computed in lattice QCD or   measured in high energy scatterings~\citep{Kharzeev:1995ij,Kharzeev:1999jt,Hatta:2018ina,Hatta:2019lxo,Wang:2019mza,Zeng:2020coc,Boussarie:2020vmu}. In particular, the tremendous progress on lattice calculations  has been made in recent years~\citep{Hagler:2003jd,Gockeler:2003jfa,Yang:2014xsa,Yang:2018nqn,Yang:2020crz}. On the other hand, the proton mass decomposition issue\citep{Lorce:2017xzd,Hatta:2018sqd,Tanaka:2018nae,Rodini:2020pis}  and the renormalization properties of the different terms of the energy momentum tensor~\citep{Hatta:2018ina,Hatta:2019lxo,Metz:2020vxd,Rodini:2020pis}, have gained  the renewed interests previously. We  recognized that the QCD EMT also plays a key role in the various different context of strongly interacting matter studies~\citep{Polyakov:2018zvc,Cosyn:2019aio,Polyakov:2019lbq,Freese:2019bhb,Sun:2020wfo,Yang:2020mtz}. In this short note, we make an attempt to promote  our understanding of the trace anomaly contribution to bound state mass with  a different strategy. Namely, we compute the contribution from the trace anomaly of the QED EMT to the mass of the simplest QED bound state, i.e.  hydrogen atom, and relate it to other known physical quantities since  this is a completely solvable  problem in quantum mechanics. 

We begin with introducing the trace of the QED EMT,
\begin{eqnarray}
    T_{\mu}^{\mu} & = & \left(1+\gamma_{m}\right)m_{0}\bar{\Psi}\Psi+\frac{\beta\left(e\right)}{2e}\left[ F^{\mu\nu}F_{\mu\nu}\right]_{R},\label{eq:EMT}
\end{eqnarray}
where  $\gamma_{m}=\frac{3\alpha_{em}}{2\pi}$ and $\frac{\beta\left(e\right)}{2e}=\frac{\alpha_{em}}{6\pi}$ at the next to leading order.  The subscript $R$ indicates that the gauge  field is the renormalized one~\citep{Adler:1976zt,Tarrach:1981bi}.  $m_0$ is electron bare mass. Notice that the first operator  is renormalization invariant~\cite{Adler:1971pm}, so that one has
$m_{0}\bar{\Psi}\Psi=\left [m\bar{\Psi}\Psi \right ]_R+O(\alpha_{em})$. 

We first compute the leading order contribution to the expectation value of the operator $m_{0}\bar{\Psi}\Psi$ for the ground state of hydrogen atom. For simplicity,  the proton inside hydrogen atom merely serves as an infinitely heavy charge source and thus  becomes decoupled in our treatment. The hydrogen atom mass at the leading order is then   given by(omitting proton mass), 
\begin{eqnarray}
M_{H,0} = \frac{\left\langle H\left|\int d^{3}x \ m_0 \bar{\Psi}(x)\Psi(x) \right|H\right\rangle}{\left\langle H| H\right\rangle}
= m \int d^3 x \ \varphi_0^\dag(x)\gamma^0 \varphi_0(x)
\end{eqnarray}
where $m_0=m$ at the leading order with $m$ being the electron physical mass.
$\varphi_0(x)$ is the ground state wave function normalized to $\int\varphi_0^{\dagger}\left(x\right)\varphi_0 \left(x\right)d^{3}x= 1$. Plugging the wave function of the ground state computed from the Dirac equation and taking the nonrelativistic limit, we derive, 
\begin{eqnarray}
    M_{H,0} = m \sqrt{1-\alpha_{em}^2}
\end{eqnarray}
The difference between $ M_{H,0}$ and electron mass is given by,
\begin{eqnarray}
    M_{H,0}-m = m \sqrt{1-\alpha_{em}^2}-m \approx -13.6 \text{eV}
\end{eqnarray}
which is  precisely the  ground state energy  of  hydrogen atom.
 Therefore, the EM potential energy and electron kinematic energy is implicitly 
 included in the electron mass term. 
 \begin{figure}[htpb]
    \includegraphics[angle=0,scale=0.5]{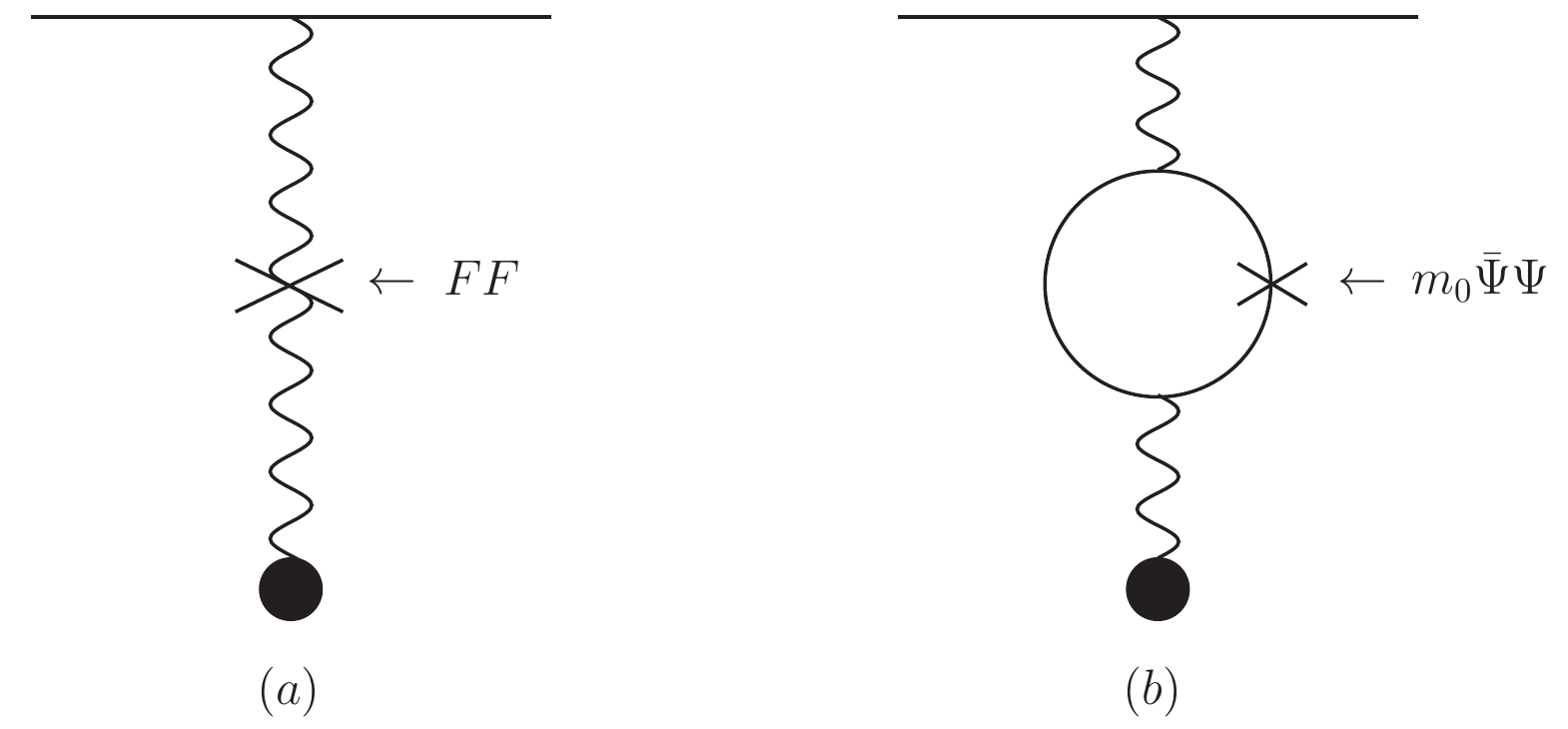}
    \caption{ Trace anomaly contribution(diagram a) and the vacuum polarization diagram with the mass operator insertion(diagram b).  Black dots represent the interaction with Coulomb potential.  } \label{vac}
  \end{figure}

 We now proceed to compute the trace anomaly contribution   and the NLO correction to the electron mass term.  At higher order, two operators $m_0 \bar{\Psi}(x)\Psi(x) $  and $F^{\mu\nu}(x)F_{\mu\nu}\left(x\right)$ mix with each other under renormalization.  As pointed out in Refs.~\citep{Metz:2020vxd,Rodini:2020pis}, the trace anomaly  contribution to the mass of the free electron state is scheme dependent, and perhaps is not a physical observable.  What matters is the difference between the anomaly contribution to the mass of free particle  and that in a bound state.  Here we specify the subtraction scheme following Ref.~\citep{Adler:1976zt,Metz:2020vxd,Rodini:2020pis}, 
 \begin{eqnarray}
  \left\langle e\left|[F^{\mu\nu}(x)F_{\mu\nu}\left(x\right)]_R \right|e\right\rangle &=&0
 \nonumber \\ 
 \left\langle \gamma \left|[F^{\mu\nu}(x)F_{\mu\nu}\left(x\right)]_R \right|\gamma \right\rangle   & = & \left\langle \gamma \left | Z_3^{-1} F^{\mu\nu}(x)F_{\mu\nu}\left(x\right) \right|\gamma \right\rangle_{\text{Tree}}
\end{eqnarray}
which appears to be the most natural choice. By inserting the time evolution operator, we calculate the expectation value of  the trace anomaly part with,
 \begin{eqnarray}
{\text Fig.~\ref{vac}(a)} = \frac{\left\langle H\left|\int d^{3}x\frac{\beta}{2e}\left [ F^{\mu\nu}(x)F_{\mu\nu}\left(x\right) \right ]_R {\cal T} e^{-i\int d^4y H_I(y)}\right|H\right\rangle}{\left\langle H| H\right\rangle}
\end{eqnarray}
The Feynman diagram  contributing to this matrix element at the first non-trivial order is shown  in Fig.\ref{vac}(a).  By applying the standard Feynman rules in the Coulomb gauge, one obtains,
\begin{eqnarray}
{\text Fig.~\ref{vac}(a)} &=&- \frac{4}{3}\alpha^2_{em} \int d^{3}y \int\frac{d^{3}  q}{(2\pi)^{3}}\frac{e^{i  \vec q\cdot \vec y}}{\vec q^{2}+i\epsilon}\left[\bar{\varphi}_0(y)\gamma^0\varphi_0\left(y\right)\right]
\end{eqnarray}
At the NLO, the expectation value of the electron mass term receives the contribution from the vacuum polarization diagram Fig.\ref{vac}(b)\cite{Cui:2011za}, 
\begin{eqnarray}
2\times {\text Fig.\ref{vac}(b)} = 8 \alpha^2_{em}\!
\int \!\!  d^{3}y \! \int\frac{d^{3}q}{(2\pi)^{3}}\frac{e^{i  \vec q\cdot \vec y}}{\vec q^{2}+i\epsilon}
\int_0^1 \!\!  da   \frac{a(1-a)m^2}{m^2+a(1-a)\vec q^2}\left[\bar{\varphi}_0(y)\gamma^0\varphi_0\left(y\right)\right]
\end{eqnarray}
 It is convenient to group these two contributions together, 
 \begin{eqnarray}
&& \!\!\!\!\!\!\!\!\!\!\!\!\!\!\!\!\!\!
{\text Fig.\ref{vac}(a)} \! +\! 2\times {\text Fig.\ref{vac}(b)} \approx 
\nonumber \\ &&  8 \alpha^2_{em} \!\!
\int \! d^{3}y \! \int\! \frac{d^{3}q}{(2\pi)^{3}}e^{i \vec q\cdot \vec y} \int_0^1 \! da 
\frac{a^2(1-a)^2}{m^2} {\varphi}_0^\dag(y)\varphi_0 \left(y\right)=\frac{-4\alpha_{em}^2}{15m^2}{\varphi}_0^\dag(0)\varphi_0\left(0\right)
\end{eqnarray}
To arrive at the above expression, we have made the Taylor expansion in terms of the power $\frac{\vec q^2}{m^2}$.  This  turns out to be just the part of the Lamb shift that is caused by the vacuum polarization effect.
\begin{figure}[htpb]
    \includegraphics[angle=0,scale=0.55]{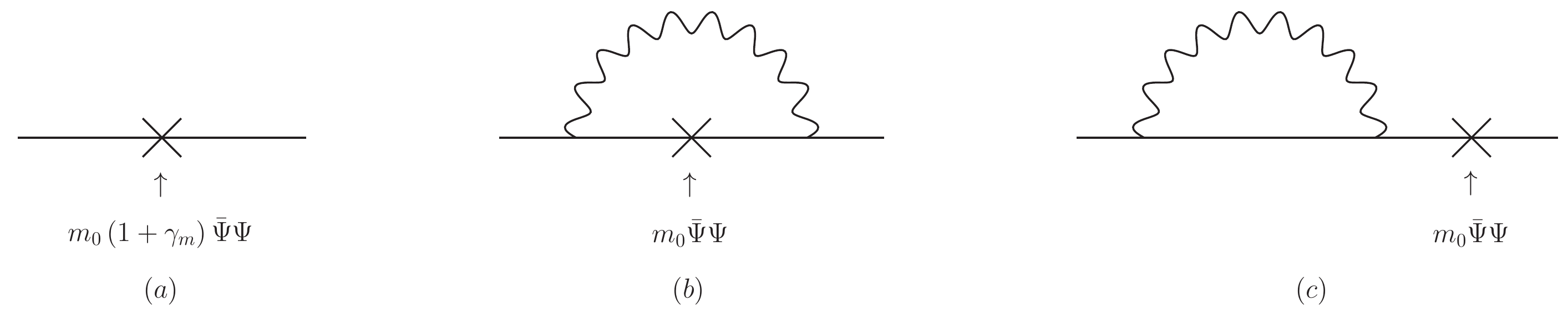}
    \caption{ NLO corrections to the electron mass term.  } \label{mass}
  \end{figure}

We now turn to the calculation of the self-energy corrections. As a warm up exercise, we first compute the self-energy corrections to the mass of free electron state as shown in Figs.~\ref{mass}. It is straightforward to obtain, 
\begin{eqnarray}
{\text Fig.\ref{mass}(a)} &=&\gamma_m m_0+m-\delta m=\gamma_m m_0+m-\frac{\alpha_{em}}{2\pi}m_0
\int_0^1 d a  (2 -a) \ln \frac{a \Lambda^2}{(1-a)^2 m_0^2}
 \\
{\text Fig.\ref{mass}(b)}&=&\frac{\alpha_{em}}{2\pi}m_0
\int_0^1 d a  \left  \{ 2\ln \frac{a \Lambda^2}{(1-a)^2 m_0^2}-\frac{2(2-a)}{(1-a) } \right \}
\\
2 \times {\text Fig.\ref{mass}(c)}&=&\frac{\alpha_{em}}{2\pi}m_0
\int_0^1 \!  d a  \left  \{ -a \ln \frac{a \Lambda^2}{(1-a)^2 m_0^2}+\frac{2a(2-a)}{(1-a) } \right \}
\end{eqnarray}
where $\Lambda$ is the UV regulator in the Pauli-Villars regularization. Summing up all three terms, one obtains,
\begin{eqnarray}
\frac{\left\langle e\left| \left(1+\gamma_{m}\right) m_0 \int d^3 x \bar{\Psi}(x)\Psi(x)\right|e\right\rangle}{\left\langle e|e\right\rangle}    = m
\label{elc}
\end{eqnarray}
As expected,  at the NLO, the physical mass of a free electron entirely comes from the electron mass term and the associated anomaly part. An all order proof of Eq.~\ref{elc}  in the subtraction scheme specified above can be achieved by invoking the Callan-Symanzik equation~\citep{Adler:1976zt}. 

Now we move on to compute the difference of the self energy correction between a free electron state and that in the bound state. The  self energy correction of electron in hydrogen atom still arises from the the loop diagrams Fig.~\ref{mass}(b) and Fig.~\ref{mass}(c), but with the internal electron propagator being replaced by the one computed in the presence of the background Coulomb field. The calculation can be most conveniently formulated using the NRQED in the Coulomb gauge.  The effective Lagrangian of NRQED is obtained from the full QED Lagrangian, 
\begin{eqnarray}
  {\cal L}  =\psi^\dag \!\left ( \! i\partial^0  -eA^0 -\frac{\vec p^2}{2m_0} +\frac{e}{2m_0}(\vec p' \!+\!\vec p)\cdot \! \vec A-\frac{e^2}{2m_0}\vec A^2 -(1+O(\alpha_{em})) \frac{ie}{2m_0} \sigma \! \cdot [ (\vec p\!-\!\vec p') \! \times \! \vec A ] \right )\! \psi+...
 \end{eqnarray}
by expressing the four component spinor $\Psi$ in terms of two component spinor $\psi$ in the nonrelativistic limit,
\begin{eqnarray}
    \Psi \approx e^{-im_0t} \frac{1}{\sqrt{2}}  \left (
\begin{array}{c}
 ( 1-\frac{\vec \sigma \cdot (\vec p-e\vec A) }{2m_0}) \psi \\( 1+\frac{\vec \sigma \cdot(\vec p-e\vec A)}{2m_0}) \psi 
 \label{NRQED}
\end{array}
\right )
\end{eqnarray}
where $\vec p$ and $\vec p'$ are three momentum operators acting on the fields $\psi$ and $\psi^\dag$ respectively. 
\begin{figure}[htpb]
    \includegraphics[angle=0,scale=0.6]{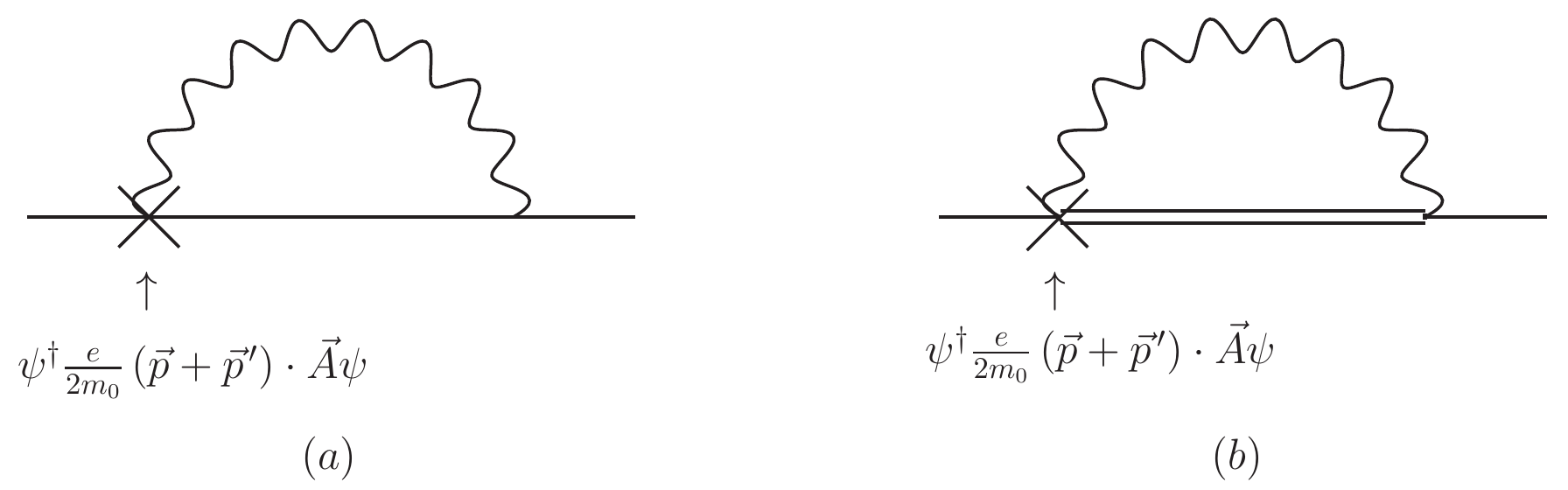}
    \caption{ Part of the NLO correction to the electron mass term  in NRQED in vacuum(diagram a), and that in the bound state(diagram b).  } \label{bound}
  \end{figure}

To simplify the calculation of Fig.\ref{mass}(c) in NRQED, we play a trick by making use of the fact that   the vector current is conserved  under renormalization, 
\begin{eqnarray}
 2\times  {\text Fig.\ref{mass}(c)}&=&-\frac{\left\langle e \left| m_0 \int d^3x  \bar{\Psi}_R(x)  \gamma_0 \Psi_R(x) \right|e \right\rangle}{\left\langle e| \int d^3x  {\Psi}_R^\dag(x)  \Psi_R(x)|e\right\rangle_{\text {Tree}}}    =  -\frac{\left\langle e \left| m_0 \int d^3x  {\Psi}_R^\dag(x)   \Psi_R(x) \right|e\right\rangle}{\left\langle e| e\right\rangle} 
\end{eqnarray}
where  $\Psi_R$  stands for the renormalized field.  Combining the above  with  Fig.\ref{mass}(b), one obtains,
\begin{eqnarray}
   {\text Fig.\ref{mass}(b)}+2\times{\text Fig.\ref{mass}(c)}=
    \frac{\left\langle e \left| m_0 \int d^3x \left [ \bar {\Psi}_R(x)   \Psi_R(x) - {\Psi}_R^\dag(x)   \Psi_R(x) \right ]\right|e\right\rangle}{\left\langle e| e\right\rangle} 
\end{eqnarray}
which can be converted into the following form with the help of Eq.~\ref{NRQED},
\begin{eqnarray}
 &&\!\!\!\!\!\!\!\!\!\!\!\!\!\!\! {\text Fig.\ref{mass}(b)}+2\times {\text Fig.\ref{mass}(c)}\approx
 \nonumber \\ && \frac{\left\langle e \left| \int d^3 x \left \{ \psi^\dag \left [  \frac{e}{2m_0}(\vec p' \!+\!\vec p)\cdot \! \vec A-\frac{\vec p^2}{2m_0}-\frac{e^2}{2m_0}\vec A^2 -\frac{ie}{2m_0} \sigma \! \cdot [ (\vec p\!-\!\vec p') \! \times \! \vec A ]\right ] \psi \right \} \right|e\right\rangle}{\left\langle e| e\right\rangle} 
\end{eqnarray}
where the operators $\psi^\dag$ and $\psi$ should be understood as the renormalized ones.  It is easy to verify that the spin-orbital coupling term $-\frac{ie}{2m_0} \sigma \! \cdot [ (\vec p\!-\!\vec p') \! \times \! \vec A ]$ does not contribute to the mass shift. We are  now ready to compute the expectation values of these operators in NRQED. 

One first notices that  the operators $ \psi^\dag\frac{\vec p^2}{2m_0} \psi $, $ \psi^\dag\frac{e^2}{2m_0}\vec A  \cdot \! \vec A  \psi $  in a free electron state and in a bound state yield the same contribution at the order of interest~\citep{Weinberg:1995mt}. The energy shift  entirely  comes from the operator  $ \psi^\dag \frac{e}{2m_0}(\vec p' \!+\!\vec p)\cdot \! \vec A \psi $. By applying the standard Feynman rules of NRQED in the Coulomb gauge, the Fig.~\ref{bound}(a) reads, 
\begin{eqnarray}
 {\text Fig.~\ref{bound}(a)}= \frac{\alpha_{em}}{(2\pi)^2} \left ( \int d^3x \varphi^\dag_0(x) \frac{\nabla^2}{m^2_0}  \varphi_0(x) \right ) \int d^3 k \left [\frac{2}{3\vec k^2} -\frac{1-\vec k^2/(3(\vec k^2+\mu^2))}{\vec k^2+\mu^2} \right ]
\end{eqnarray}
To arrive at the above expression, we have imposed a scale cut off $\mu$ by replacing the photon propagator $\frac{1}{k^2+i\epsilon} \rightarrow \frac{1}{k^2+i\epsilon}-\frac{1}{k^2-\mu^2+i\epsilon}$ following the Weinberg's method~\citep{Weinberg:1995mt}. In a bound state, the internal electron propagator in the position space is given by $\sum_M \frac{\varphi_M(x) \varphi_M^\dag(y)}{\Delta E_M+i\epsilon}$, where $\varphi_M$ are a complete orthonormal set of state-vectors, and $\Delta E_M$ is  the difference between the energies of the  state  M and the ground state.  Here we neglect the positron contribution to the propagator. Using the electron propagator in the bound state, one readily obtains, 
\begin{eqnarray}
 {\text Fig.\ref{bound}(b)}&= &\frac{\alpha_{em}}{(2\pi)^2} \sum_M \left | \int d^3x \varphi^\dag_0(x) \frac{\nabla}{m_0}  \varphi_M(x) \right |^2 \nonumber \\
    &&\times \int d^3 k \left [\frac{2}{3|\vec k | (\Delta E_M+|\vec k|)} -\frac{1-\vec k^2/(3(\vec k^2+\mu^2))}{\sqrt {\vec k^2+\mu^2} (\Delta E_M+\sqrt {\vec k^2+\mu^2})} \right ]
 \end{eqnarray}
 The energy shift for the ground state is then given by,
 \begin{eqnarray}
  {\text Fig.\ref{bound}(b)}-{\text Fig.\ref{bound}(a)} \approx \frac{4\alpha^2_{em}}{3m^2} \varphi^\dag_0(0)\varphi_0(0) \left [  \ln \frac{\mu}{2\Delta E}+\frac{5}{6} \right ]
\end{eqnarray}
where $\Delta E$ is a mean excitation energy. This result  was first derived by Bethe in 1947~\cite{Bethe:1947id}.  To justify the nonrelativistic treatment for electron, the scale $\mu$ should be chosen to be smaller than $m$, and was actually just set to be $m$ in the Bethe's original work, leading to a rather good agreement with the observed energy shift.  To remove the $\mu$ dependence in a rigorous way, one has to include the high energy part~\citep{Weinberg:1995mt}, or in  modern effective field language, performs a matching calculation for the vertex correction between the full QED and NRQED~\citep{Labelle:1996uc,Porto:2017shd,Huang:2019hdj}, which is however  beyond the focus  of this short note. 

In summary, we have calculated the contribution from the trace anomaly of the QED energy momentum tensor to the hydrogen atom mass(Eq.9).  It is shown to be related to the small part of the Lamb shift splitting that arises from the vacuum polarization effect.  We further computed the radiative corrections to the electron mass term of the QED EMT in  the bound state(Eq.21), which turns out to be consistent with the electron  self-energy calculation in the conventional  treatment of the Lamb shift problem, as it should be.   The trace anomaly part  is widely believed to be the dominant contribution to the mass of QCD bound states. It would be interesting  to extend this  analysis  to heavy quarkonium system with an effective potential model and compare with the Lattice result~\cite{Sun:2020pda} in the future. 

{ \bf Acknowledgments:}
    J. Zhou thanks Yi-bo Yang and Jian-hua Gao for helpful discussions. Bao-dong Sun Thanks Ren-hong Fang and Rui Yu for helpful discussions. J. Zhou has been supported by the National Nature Science Foundations of China under Grant No.\ 11675093. B.-D. Sun has been supported by the National Natural Science Foundation of China under Grant No. 11947228 and the China Postdoctoral Science Foundation under Grant No. 2019M662316.

\bibliography{ref.bib}

\begin{thebibliography}{38}
\expandafter\ifx\csname natexlab\endcsname\relax\def\natexlab#1{#1}\fi
\expandafter\ifx\csname bibnamefont\endcsname\relax
  \def\bibnamefont#1{#1}\fi
\expandafter\ifx\csname bibfnamefont\endcsname\relax
  \def\bibfnamefont#1{#1}\fi
\expandafter\ifx\csname citenamefont\endcsname\relax
  \def\citenamefont#1{#1}\fi
\expandafter\ifx\csname url\endcsname\relax
  \def\url#1{\texttt{#1}}\fi
\expandafter\ifx\csname urlprefix\endcsname\relax\def\urlprefix{URL }\fi
\providecommand{\bibinfo}[2]{#2}
\providecommand{\eprint}[2][]{\url{#2}}

\bibitem[{\citenamefont{Brack}(1983)}]{Brack:1983ht}
\bibinfo{author}{\bibfnamefont{M.}~\bibnamefont{Brack}},
  \bibinfo{journal}{Phys. Rev. D} \textbf{\bibinfo{volume}{27}},
  \bibinfo{pages}{1950} (\bibinfo{year}{1983}).

\bibitem[{\citenamefont{Adler et~al.}(1977)\citenamefont{Adler, Collins, and
  Duncan}}]{Adler:1976zt}
\bibinfo{author}{\bibfnamefont{S.~L.} \bibnamefont{Adler}},
  \bibinfo{author}{\bibfnamefont{J.~C.} \bibnamefont{Collins}},
  \bibnamefont{and} \bibinfo{author}{\bibfnamefont{A.}~\bibnamefont{Duncan}},
  \bibinfo{journal}{Phys. Rev. D} \textbf{\bibinfo{volume}{15}},
  \bibinfo{pages}{1712} (\bibinfo{year}{1977}).

\bibitem[{\citenamefont{Nielsen}(1977)}]{Nielsen:1977sy}
\bibinfo{author}{\bibfnamefont{N.}~\bibnamefont{Nielsen}},
  \bibinfo{journal}{Nucl. Phys. B} \textbf{\bibinfo{volume}{120}},
  \bibinfo{pages}{212} (\bibinfo{year}{1977}).

\bibitem[{\citenamefont{Collins et~al.}(1977)\citenamefont{Collins, Duncan, and
  Joglekar}}]{Collins:1976yq}
\bibinfo{author}{\bibfnamefont{J.~C.} \bibnamefont{Collins}},
  \bibinfo{author}{\bibfnamefont{A.}~\bibnamefont{Duncan}}, \bibnamefont{and}
  \bibinfo{author}{\bibfnamefont{S.~D.} \bibnamefont{Joglekar}},
  \bibinfo{journal}{Phys. Rev. D} \textbf{\bibinfo{volume}{16}},
  \bibinfo{pages}{438} (\bibinfo{year}{1977}).

\bibitem[{\citenamefont{Ji}(1995{\natexlab{a}})}]{Ji:1994av}
\bibinfo{author}{\bibfnamefont{X.-D.} \bibnamefont{Ji}},
  \bibinfo{journal}{Phys. Rev. Lett.} \textbf{\bibinfo{volume}{74}},
  \bibinfo{pages}{1071} (\bibinfo{year}{1995}{\natexlab{a}}),
  \eprint{hep-ph/9410274}.

\bibitem[{\citenamefont{Ji}(1995{\natexlab{b}})}]{Ji:1995sv}
\bibinfo{author}{\bibfnamefont{X.-D.} \bibnamefont{Ji}},
  \bibinfo{journal}{Phys. Rev. D} \textbf{\bibinfo{volume}{52}},
  \bibinfo{pages}{271} (\bibinfo{year}{1995}{\natexlab{b}}),
  \eprint{hep-ph/9502213}.

\bibitem[{\citenamefont{Kharzeev}(1996)}]{Kharzeev:1995ij}
\bibinfo{author}{\bibfnamefont{D.}~\bibnamefont{Kharzeev}},
  \bibinfo{journal}{Proc. Int. Sch. Phys. Fermi}
  \textbf{\bibinfo{volume}{130}}, \bibinfo{pages}{105} (\bibinfo{year}{1996}),
  \eprint{nucl-th/9601029}.

\bibitem[{\citenamefont{Kharzeev et~al.}(1999)\citenamefont{Kharzeev, Satz,
  Syamtomov, and Zinovev}}]{Kharzeev:1999jt}
\bibinfo{author}{\bibfnamefont{D.}~\bibnamefont{Kharzeev}},
  \bibinfo{author}{\bibfnamefont{H.}~\bibnamefont{Satz}},
  \bibinfo{author}{\bibfnamefont{A.}~\bibnamefont{Syamtomov}},
  \bibnamefont{and} \bibinfo{author}{\bibfnamefont{G.}~\bibnamefont{Zinovev}},
  \bibinfo{journal}{Nucl. Phys. A} \textbf{\bibinfo{volume}{661}},
  \bibinfo{pages}{568} (\bibinfo{year}{1999}).

\bibitem[{\citenamefont{Hatta and Yang}(2018)}]{Hatta:2018ina}
\bibinfo{author}{\bibfnamefont{Y.}~\bibnamefont{Hatta}} \bibnamefont{and}
  \bibinfo{author}{\bibfnamefont{D.-L.} \bibnamefont{Yang}},
  \bibinfo{journal}{Phys. Rev. D} \textbf{\bibinfo{volume}{98}},
  \bibinfo{pages}{074003} (\bibinfo{year}{2018}), \eprint{1808.02163}.

\bibitem[{\citenamefont{Hatta et~al.}(2019)\citenamefont{Hatta, Rajan, and
  Yang}}]{Hatta:2019lxo}
\bibinfo{author}{\bibfnamefont{Y.}~\bibnamefont{Hatta}},
  \bibinfo{author}{\bibfnamefont{A.}~\bibnamefont{Rajan}}, \bibnamefont{and}
  \bibinfo{author}{\bibfnamefont{D.-L.} \bibnamefont{Yang}},
  \bibinfo{journal}{Phys. Rev. D} \textbf{\bibinfo{volume}{100}},
  \bibinfo{pages}{014032} (\bibinfo{year}{2019}), \eprint{1906.00894}.

\bibitem[{\citenamefont{Wang et~al.}(2020)\citenamefont{Wang, Evslin, and
  Chen}}]{Wang:2019mza}
\bibinfo{author}{\bibfnamefont{R.}~\bibnamefont{Wang}},
  \bibinfo{author}{\bibfnamefont{J.}~\bibnamefont{Evslin}}, \bibnamefont{and}
  \bibinfo{author}{\bibfnamefont{X.}~\bibnamefont{Chen}},
  \bibinfo{journal}{Eur. Phys. J. C} \textbf{\bibinfo{volume}{80}},
  \bibinfo{pages}{507} (\bibinfo{year}{2020}), \eprint{1912.12040}.

\bibitem[{\citenamefont{Zeng et~al.}(2020)\citenamefont{Zeng, Wang, Zhang, Xie,
  Wang, and Chen}}]{Zeng:2020coc}
\bibinfo{author}{\bibfnamefont{F.}~\bibnamefont{Zeng}},
  \bibinfo{author}{\bibfnamefont{X.-Y.} \bibnamefont{Wang}},
  \bibinfo{author}{\bibfnamefont{L.}~\bibnamefont{Zhang}},
  \bibinfo{author}{\bibfnamefont{Y.-P.} \bibnamefont{Xie}},
  \bibinfo{author}{\bibfnamefont{R.}~\bibnamefont{Wang}}, \bibnamefont{and}
  \bibinfo{author}{\bibfnamefont{X.}~\bibnamefont{Chen}}
  (\bibinfo{year}{2020}), \eprint{2008.13439}.

\bibitem[{\citenamefont{Boussarie and Hatta}(2020)}]{Boussarie:2020vmu}
\bibinfo{author}{\bibfnamefont{R.}~\bibnamefont{Boussarie}} \bibnamefont{and}
  \bibinfo{author}{\bibfnamefont{Y.}~\bibnamefont{Hatta}},
  \bibinfo{journal}{Phys. Rev. D} \textbf{\bibinfo{volume}{101}},
  \bibinfo{pages}{114004} (\bibinfo{year}{2020}), \eprint{2004.12715}.

\bibitem[{\citenamefont{Hagler et~al.}(2003)\citenamefont{Hagler, Negele,
  Renner, Schroers, Lippert, and Schilling}}]{Hagler:2003jd}
\bibinfo{author}{\bibfnamefont{P.}~\bibnamefont{Hagler}},
  \bibinfo{author}{\bibfnamefont{J.~W.} \bibnamefont{Negele}},
  \bibinfo{author}{\bibfnamefont{D.~B.} \bibnamefont{Renner}},
  \bibinfo{author}{\bibfnamefont{W.}~\bibnamefont{Schroers}},
  \bibinfo{author}{\bibfnamefont{T.}~\bibnamefont{Lippert}}, \bibnamefont{and}
  \bibinfo{author}{\bibfnamefont{K.}~\bibnamefont{Schilling}}
  (\bibinfo{collaboration}{LHPC, SESAM}), \bibinfo{journal}{Phys. Rev. D}
  \textbf{\bibinfo{volume}{68}}, \bibinfo{pages}{034505}
  (\bibinfo{year}{2003}), \eprint{hep-lat/0304018}.

\bibitem[{\citenamefont{Gockeler et~al.}(2004)\citenamefont{Gockeler, Horsley,
  Pleiter, Rakow, Schafer, Schierholz, and Schroers}}]{Gockeler:2003jfa}
\bibinfo{author}{\bibfnamefont{M.}~\bibnamefont{Gockeler}},
  \bibinfo{author}{\bibfnamefont{R.}~\bibnamefont{Horsley}},
  \bibinfo{author}{\bibfnamefont{D.}~\bibnamefont{Pleiter}},
  \bibinfo{author}{\bibfnamefont{P.~E.} \bibnamefont{Rakow}},
  \bibinfo{author}{\bibfnamefont{A.}~\bibnamefont{Schafer}},
  \bibinfo{author}{\bibfnamefont{G.}~\bibnamefont{Schierholz}},
  \bibnamefont{and} \bibinfo{author}{\bibfnamefont{W.}~\bibnamefont{Schroers}}
  (\bibinfo{collaboration}{QCDSF}), \bibinfo{journal}{Phys. Rev. Lett.}
  \textbf{\bibinfo{volume}{92}}, \bibinfo{pages}{042002}
  (\bibinfo{year}{2004}), \eprint{hep-ph/0304249}.

\bibitem[{\citenamefont{Yang et~al.}(2015)\citenamefont{Yang, Chen, Draper,
  Gong, Liu, Liu, and Ma}}]{Yang:2014xsa}
\bibinfo{author}{\bibfnamefont{Y.-B.} \bibnamefont{Yang}},
  \bibinfo{author}{\bibfnamefont{Y.}~\bibnamefont{Chen}},
  \bibinfo{author}{\bibfnamefont{T.}~\bibnamefont{Draper}},
  \bibinfo{author}{\bibfnamefont{M.}~\bibnamefont{Gong}},
  \bibinfo{author}{\bibfnamefont{K.-F.} \bibnamefont{Liu}},
  \bibinfo{author}{\bibfnamefont{Z.}~\bibnamefont{Liu}}, \bibnamefont{and}
  \bibinfo{author}{\bibfnamefont{J.-P.} \bibnamefont{Ma}},
  \bibinfo{journal}{Phys. Rev. D} \textbf{\bibinfo{volume}{91}},
  \bibinfo{pages}{074516} (\bibinfo{year}{2015}), \eprint{1405.4440}.

\bibitem[{\citenamefont{Yang et~al.}(2018)\citenamefont{Yang, Liang, Bi, Chen,
  Draper, Liu, and Liu}}]{Yang:2018nqn}
\bibinfo{author}{\bibfnamefont{Y.-B.} \bibnamefont{Yang}},
  \bibinfo{author}{\bibfnamefont{J.}~\bibnamefont{Liang}},
  \bibinfo{author}{\bibfnamefont{Y.-J.} \bibnamefont{Bi}},
  \bibinfo{author}{\bibfnamefont{Y.}~\bibnamefont{Chen}},
  \bibinfo{author}{\bibfnamefont{T.}~\bibnamefont{Draper}},
  \bibinfo{author}{\bibfnamefont{K.-F.} \bibnamefont{Liu}}, \bibnamefont{and}
  \bibinfo{author}{\bibfnamefont{Z.}~\bibnamefont{Liu}},
  \bibinfo{journal}{Phys. Rev. Lett.} \textbf{\bibinfo{volume}{121}},
  \bibinfo{pages}{212001} (\bibinfo{year}{2018}), \eprint{1808.08677}.

\bibitem[{\citenamefont{Yang et~al.}(2020{\natexlab{a}})\citenamefont{Yang,
  Liang, Liu, and Sun}}]{Yang:2020crz}
\bibinfo{author}{\bibfnamefont{Y.-B.} \bibnamefont{Yang}},
  \bibinfo{author}{\bibfnamefont{J.}~\bibnamefont{Liang}},
  \bibinfo{author}{\bibfnamefont{Z.}~\bibnamefont{Liu}}, \bibnamefont{and}
  \bibinfo{author}{\bibfnamefont{P.}~\bibnamefont{Sun}}
  (\bibinfo{collaboration}{xQCD}), \bibinfo{journal}{PoS}
  \textbf{\bibinfo{volume}{LATTICE2019}}, \bibinfo{pages}{001}
  (\bibinfo{year}{2020}{\natexlab{a}}), \eprint{2003.12914}.

\bibitem[{\citenamefont{Lorce}(2018)}]{Lorce:2017xzd}
\bibinfo{author}{\bibfnamefont{C.}~\bibnamefont{Lorce}}, \bibinfo{journal}{Eur.
  Phys. J. C} \textbf{\bibinfo{volume}{78}}, \bibinfo{pages}{120}
  (\bibinfo{year}{2018}), \eprint{1706.05853}.

\bibitem[{\citenamefont{Hatta et~al.}(2018)\citenamefont{Hatta, Rajan, and
  Tanaka}}]{Hatta:2018sqd}
\bibinfo{author}{\bibfnamefont{Y.}~\bibnamefont{Hatta}},
  \bibinfo{author}{\bibfnamefont{A.}~\bibnamefont{Rajan}}, \bibnamefont{and}
  \bibinfo{author}{\bibfnamefont{K.}~\bibnamefont{Tanaka}},
  \bibinfo{journal}{JHEP} \textbf{\bibinfo{volume}{12}}, \bibinfo{pages}{008}
  (\bibinfo{year}{2018}), \eprint{1810.05116}.

\bibitem[{\citenamefont{Tanaka}(2019)}]{Tanaka:2018nae}
\bibinfo{author}{\bibfnamefont{K.}~\bibnamefont{Tanaka}},
  \bibinfo{journal}{JHEP} \textbf{\bibinfo{volume}{01}}, \bibinfo{pages}{120}
  (\bibinfo{year}{2019}), \eprint{1811.07879}.

\bibitem[{\citenamefont{Rodini et~al.}(2020)\citenamefont{Rodini, Metz, and
  Pasquini}}]{Rodini:2020pis}
\bibinfo{author}{\bibfnamefont{S.}~\bibnamefont{Rodini}},
  \bibinfo{author}{\bibfnamefont{A.}~\bibnamefont{Metz}}, \bibnamefont{and}
  \bibinfo{author}{\bibfnamefont{B.}~\bibnamefont{Pasquini}},
  \bibinfo{journal}{JHEP} \textbf{\bibinfo{volume}{09}}, \bibinfo{pages}{067}
  (\bibinfo{year}{2020}), \eprint{2004.03704}.

\bibitem[{\citenamefont{Metz et~al.}(2020)\citenamefont{Metz, Pasquini, and
  Rodini}}]{Metz:2020vxd}
\bibinfo{author}{\bibfnamefont{A.}~\bibnamefont{Metz}},
  \bibinfo{author}{\bibfnamefont{B.}~\bibnamefont{Pasquini}}, \bibnamefont{and}
  \bibinfo{author}{\bibfnamefont{S.}~\bibnamefont{Rodini}},
  \bibinfo{journal}{arXiv:2006.11171}  (\bibinfo{year}{2020}).

\bibitem[{\citenamefont{Polyakov and Schweitzer}(2018)}]{Polyakov:2018zvc}
\bibinfo{author}{\bibfnamefont{M.~V.} \bibnamefont{Polyakov}} \bibnamefont{and}
  \bibinfo{author}{\bibfnamefont{P.}~\bibnamefont{Schweitzer}},
  \bibinfo{journal}{Int. J. Mod. Phys. A} \textbf{\bibinfo{volume}{33}},
  \bibinfo{pages}{1830025} (\bibinfo{year}{2018}), \eprint{1805.06596}.

\bibitem[{\citenamefont{Cosyn et~al.}(2019)\citenamefont{Cosyn, Cotogno,
  Freese, and Lorcé}}]{Cosyn:2019aio}
\bibinfo{author}{\bibfnamefont{W.}~\bibnamefont{Cosyn}},
  \bibinfo{author}{\bibfnamefont{S.}~\bibnamefont{Cotogno}},
  \bibinfo{author}{\bibfnamefont{A.}~\bibnamefont{Freese}}, \bibnamefont{and}
  \bibinfo{author}{\bibfnamefont{C.}~\bibnamefont{Lorcé}},
  \bibinfo{journal}{Eur. Phys. J. C} \textbf{\bibinfo{volume}{79}},
  \bibinfo{pages}{476} (\bibinfo{year}{2019}), \eprint{1903.00408}.

\bibitem[{\citenamefont{Polyakov and Sun}(2019)}]{Polyakov:2019lbq}
\bibinfo{author}{\bibfnamefont{M.~V.} \bibnamefont{Polyakov}} \bibnamefont{and}
  \bibinfo{author}{\bibfnamefont{B.-D.} \bibnamefont{Sun}},
  \bibinfo{journal}{Phys. Rev. D} \textbf{\bibinfo{volume}{100}},
  \bibinfo{pages}{036003} (\bibinfo{year}{2019}), \eprint{1903.02738}.

\bibitem[{\citenamefont{Freese and Cloët}(2019)}]{Freese:2019bhb}
\bibinfo{author}{\bibfnamefont{A.}~\bibnamefont{Freese}} \bibnamefont{and}
  \bibinfo{author}{\bibfnamefont{I.~C.} \bibnamefont{Cloët}},
  \bibinfo{journal}{Phys. Rev. C} \textbf{\bibinfo{volume}{100}},
  \bibinfo{pages}{015201} (\bibinfo{year}{2019}), \eprint{1903.09222}.

\bibitem[{\citenamefont{Sun and Dong}(2020)}]{Sun:2020wfo}
\bibinfo{author}{\bibfnamefont{B.-D.} \bibnamefont{Sun}} \bibnamefont{and}
  \bibinfo{author}{\bibfnamefont{Y.-B.} \bibnamefont{Dong}},
  \bibinfo{journal}{Phys. Rev. D} \textbf{\bibinfo{volume}{101}},
  \bibinfo{pages}{096008} (\bibinfo{year}{2020}), \eprint{2002.02648}.

\bibitem[{\citenamefont{Yang et~al.}(2020{\natexlab{b}})\citenamefont{Yang,
  Gao, Liang, and Wang}}]{Yang:2020mtz}
\bibinfo{author}{\bibfnamefont{S.-Z.} \bibnamefont{Yang}},
  \bibinfo{author}{\bibfnamefont{J.-H.} \bibnamefont{Gao}},
  \bibinfo{author}{\bibfnamefont{Z.-T.} \bibnamefont{Liang}}, \bibnamefont{and}
  \bibinfo{author}{\bibfnamefont{Q.}~\bibnamefont{Wang}}
  (\bibinfo{year}{2020}{\natexlab{b}}), \eprint{2003.04517}.

\bibitem[{\citenamefont{Tarrach}(1982)}]{Tarrach:1981bi}
\bibinfo{author}{\bibfnamefont{R.}~\bibnamefont{Tarrach}},
  \bibinfo{journal}{Nucl. Phys. B} \textbf{\bibinfo{volume}{196}},
  \bibinfo{pages}{45} (\bibinfo{year}{1982}).

\bibitem[{\citenamefont{Adler and Bardeen}(1971)}]{Adler:1971pm}
\bibinfo{author}{\bibfnamefont{S.~L.} \bibnamefont{Adler}} \bibnamefont{and}
  \bibinfo{author}{\bibfnamefont{W.~A.} \bibnamefont{Bardeen}},
  \bibinfo{journal}{Phys. Rev. D} \textbf{\bibinfo{volume}{4}},
  \bibinfo{pages}{3045} (\bibinfo{year}{1971}), \bibinfo{note}{[Erratum:
  Phys.Rev.D 6, 734 (1972)]}.

\bibitem[{\citenamefont{Cui et~al.}(2011)\citenamefont{Cui, Ma, and
  Wu}}]{Cui:2011za}
\bibinfo{author}{\bibfnamefont{J.-W.} \bibnamefont{Cui}},
  \bibinfo{author}{\bibfnamefont{Y.-L.} \bibnamefont{Ma}}, \bibnamefont{and}
  \bibinfo{author}{\bibfnamefont{Y.-L.} \bibnamefont{Wu}},
  \bibinfo{journal}{Phys. Rev. D} \textbf{\bibinfo{volume}{84}},
  \bibinfo{pages}{025020} (\bibinfo{year}{2011}), \eprint{1103.2026}.

\bibitem[{\citenamefont{Weinberg}(2005)}]{Weinberg:1995mt}
\bibinfo{author}{\bibfnamefont{S.}~\bibnamefont{Weinberg}},
  \emph{\bibinfo{title}{{The Quantum theory of fields. Vol. 1: Foundations}}}
  (\bibinfo{publisher}{Cambridge University Press}, \bibinfo{year}{2005}), ISBN
  \bibinfo{isbn}{978-0-521-67053-1, 978-0-511-25204-4}.

\bibitem[{\citenamefont{Bethe}(1947)}]{Bethe:1947id}
\bibinfo{author}{\bibfnamefont{H.}~\bibnamefont{Bethe}},
  \bibinfo{journal}{Phys. Rev.} \textbf{\bibinfo{volume}{72}},
  \bibinfo{pages}{339} (\bibinfo{year}{1947}).

\bibitem[{\citenamefont{Labelle and Mohammad~Zebarjad}(1999)}]{Labelle:1996uc}
\bibinfo{author}{\bibfnamefont{P.}~\bibnamefont{Labelle}} \bibnamefont{and}
  \bibinfo{author}{\bibfnamefont{S.}~\bibnamefont{Mohammad~Zebarjad}},
  \bibinfo{journal}{Can. J. Phys.} \textbf{\bibinfo{volume}{77}},
  \bibinfo{pages}{267} (\bibinfo{year}{1999}), \eprint{hep-ph/9611313}.

\bibitem[{\citenamefont{Porto}(2017)}]{Porto:2017shd}
\bibinfo{author}{\bibfnamefont{R.~A.} \bibnamefont{Porto}},
  \bibinfo{journal}{Phys. Rev. D} \textbf{\bibinfo{volume}{96}},
  \bibinfo{pages}{024063} (\bibinfo{year}{2017}), \eprint{1703.06434}.

\bibitem[{\citenamefont{Huang et~al.}(2019)\citenamefont{Huang, Jia, and
  Yu}}]{Huang:2019hdj}
\bibinfo{author}{\bibfnamefont{Y.}~\bibnamefont{Huang}},
  \bibinfo{author}{\bibfnamefont{Y.}~\bibnamefont{Jia}}, \bibnamefont{and}
  \bibinfo{author}{\bibfnamefont{R.}~\bibnamefont{Yu}} (\bibinfo{year}{2019}),
  \eprint{1901.04971}.

\bibitem[{\citenamefont{Sun et~al.}(2020)\citenamefont{Sun, Chen, Sun, and
  Yang}}]{Sun:2020pda}
\bibinfo{author}{\bibfnamefont{W.}~\bibnamefont{Sun}},
  \bibinfo{author}{\bibfnamefont{Y.}~\bibnamefont{Chen}},
  \bibinfo{author}{\bibfnamefont{P.}~\bibnamefont{Sun}}, \bibnamefont{and}
  \bibinfo{author}{\bibfnamefont{Y.-B.} \bibnamefont{Yang}}
  (\bibinfo{collaboration}{\ensuremath{\chi}QCD}) (\bibinfo{year}{2020}),
  \eprint{2012.06228}.

\end{thebibliography}

\end{document}